\documentclass{article}
\usepackage[accepted]{icml2022preprint}
\usepackage{hyperref}

\usepackage{soul}
\usepackage{url}
\usepackage[utf8]{inputenc}
\usepackage{caption}
\usepackage{graphicx}   
\usepackage{amsmath}
\usepackage{xcolor}
\usepackage{amssymb}
\usepackage{bm}
\usepackage{booktabs}
\usepackage{tikz}
\DeclareMathOperator{\softmax}{softmax}

\definecolor{powerpointorange}{RGB}{184,96,41}
\definecolor{powerpointgreen}{RGB}{94,128,63}
\definecolor{powerpointblue}{RGB}{64,117,177}
 \author{
Michael J. Curry$^1$\and
Tuomas Sandholm$^{2,3}$\and
John Dickerson$^1$\\
\affiliations
$^1$University of Maryland\\
$^2$Carnegie Mellon University\\
$^3$Optimized Markets, Inc., Strategic Machine, Inc. Strategy Robot, Inc.\\
\emails
curry@cs.umd.edu, sandholm@cs.cmu.edu, johnd@umd.edu
}

\begin{document}
\twocolumn[\icmltitle{Differentiable Economics for Randomized Affine Maximizer Auctions}
\begin{icmlauthorlist}
\icmlauthor{Michael Curry}{umd}
\icmlauthor{Tuomas Sandholm}{cmu,optimizedmarkets,strategicmachine,strategyrobot}
\icmlauthor{John Dickerson}{umd}
\end{icmlauthorlist}

\icmlaffiliation{umd}{University of Maryland}
\icmlaffiliation{cmu}{Carnegie Mellon University}
\icmlaffiliation{optimizedmarkets}{Optimized Markets, Inc.}
\icmlaffiliation{strategicmachine}{Strategic Machine, Inc.}
\icmlaffiliation{strategyrobot}{Strategy Robot, Inc.}

\icmlcorrespondingauthor{Michael Curry}{curry@cs.umd.edu}
\vskip 0.3in
]
\printAffiliationsAndNotice{}
\begin{abstract}
    A recent approach to automated mechanism design, differentiable economics, represents auctions by rich function approximators and optimizes their performance by gradient descent. The ideal auction architecture for differentiable economics would be perfectly strategyproof, support multiple bidders and items, and be rich enough to represent the optimal (i.e. revenue-maximizing) mechanism. So far, such an architecture does not exist. There are single-bidder approaches (MenuNet, RochetNet) which are always strategyproof and can represent optimal mechanisms. RegretNet is multi-bidder and can approximate any mechanism, but is only approximately strategyproof. We present an architecture that supports multiple bidders and is perfectly strategyproof, but cannot necessarily represent the optimal mechanism. This architecture is the classic affine maximizer auction (AMA), modified to offer lotteries. By using the gradient-based optimization tools of differentiable economics, we can now train lottery AMAs, competing with or outperforming prior approaches in revenue.
\end{abstract}

\section{Introduction}

\begin{figure}[t]
    \centering
    \includegraphics[width=0.6\columnwidth]{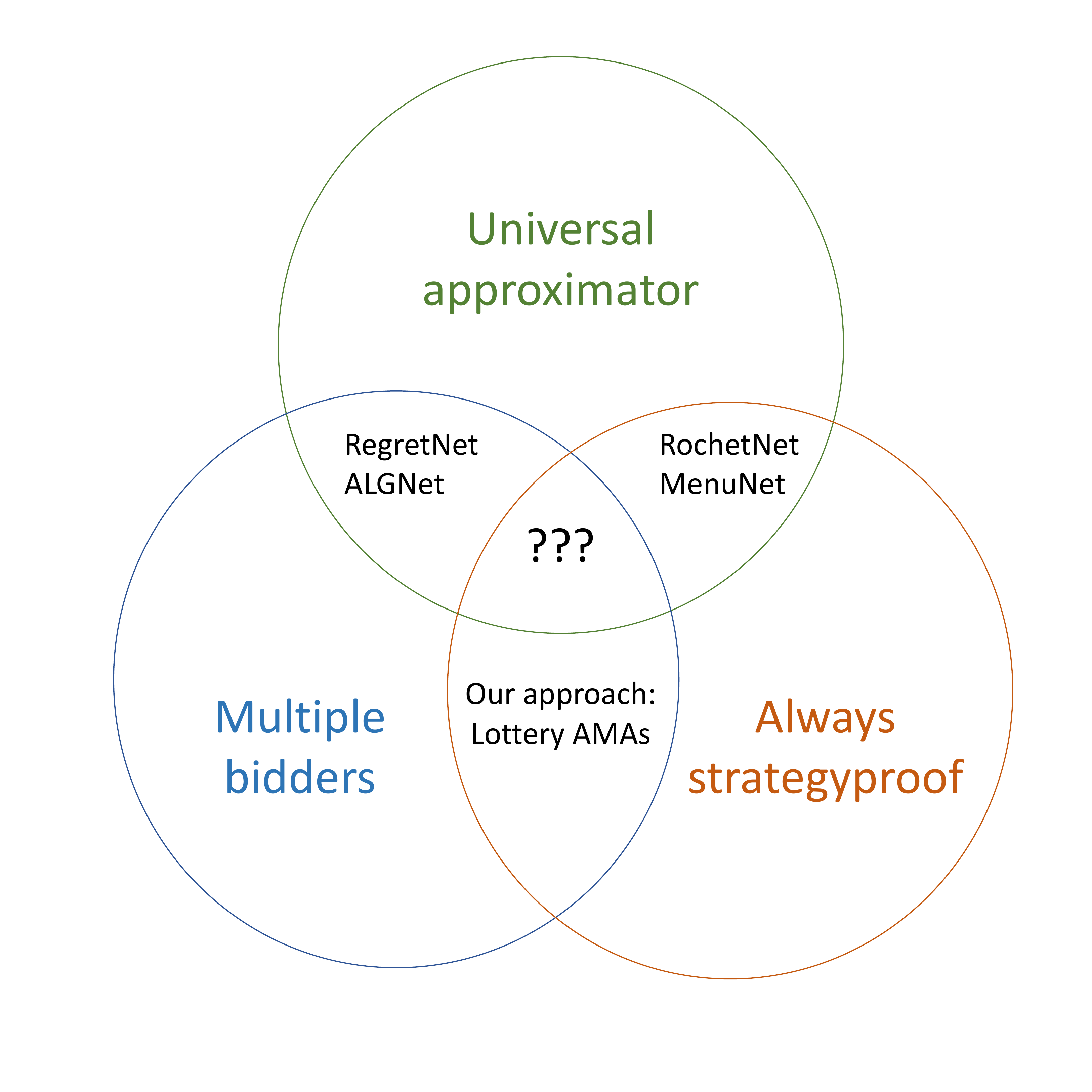}
    \vspace{-.2in}
    \caption{Our architecture in relation to other techniques from differentiable economics for multi-item revenue-maximizing auction design. The ``holy grail'' in the middle of the Venn diagram---that is, techniques that can represent \textcolor{powerpointgreen}{(i) any auction} for \textcolor{powerpointblue}{(ii) general numbers of bidders and items} while
    \textcolor{powerpointorange}{(iii) guaranteeing strategyproofness}---has not been achieved; however, we show that our method achieves (iii) strategyproofness-by-design for (ii) general numbers of items and bidders while still improving revenue over baselines.}
    \label{fig:ama_venn}
\end{figure}

Auctions are a widely-used mechanism for allocating scarce items that are for sale, in which a centralized auctioneer solicits bids from auction participants, and based on those bids, allocates the items (possibly keeping some of them) and charges some payments. 
The auctioneer may wish to design the auction to achieve some goal. The usual assumption is that the auctioneer has access to a prior distribution over bidders' valuations.
Typically, it is also desired that the auction be strategyproof, that is, there should be no incentive for bidders to be untruthful in their bids about their valuations.

When the auctioneer wants to maximize the total welfare of the bidders, the \textit{Vickrey-Clarke-Groves (VCG)} mechanism, which is always strategyproof, is also optimal~\cite{Vickrey1961Counterspeculation,Clarke1971Multipart,Groves1973Incentives}.
When the auctioneer instead wants to maximize her revenue (or profit), the problem is significantly  more challenging.

\citet{Myerson1981Optimal} settled the revenue-maximizing strategyproof auction problem when there is one item for sale. \citet{Maskin89Optimal} generalized that mechanism to the case of multiple copies of a single item. However, four decades later, the multi-item revenue-maximizing auction is still unknown. Special cases of the two-item setting have been solved~\cite{Armstrong00Optimal,Avery00Bundling}. 
There is some theory of strong duality \cite{Daskalakis2017Strong,Kash2016Optimal} for selling multiple items to a single agent. 
There have also been some successes for the weaker notion of Bayesian incentive compatibility~\cite{Cai2012algorithmic,Cai2012Optimal,Cai2013Understanding}.
But for designing dominant-strategy incentive compatible mechanisms that sell multiple items to multiple agents there has been little progress despite decades of research.
\citet{Yao2017Dominant} presents a result for one special case, giving an explicit example of a revenue gap between the best dominant-strategy incentive compatible mechanism and the best Bayes-Nash incentive compatible mechanism.
Nevertheless, the problem is wide open. Even for the seemingly trivial case of two agents with i.i.d. uniform valuations over two items, the optimal selling mechanism is not known.

In part motivated by the fact that the theory on this question has essentially gotten stuck for decades, \citet{Conitzer02Mechanism,Sandholm2003Automated} introduced the idea of \textit{automated mechanism design (AMD)}: designing the mechanism computationally for the problem instance at hand, as opposed to trying to analytically derive a general form for the revenue-maximizing multi-item auction. AMD has since become a popular research topic. Three different high-level approaches to AMD have been introduced: 1) designing the mechanism from scratch in tabular form~\citep{Conitzer02Mechanism}, 2) conducting search over the parameters of a mechanism class where all the mechanisms in the class have some desirable properties such as strategyproofness and individual rationality (the latter incentivizes buyers to participate)~\cite{Likhodedov2004Methods, Likhodedov2005Approximating,Sandholm2015Automated}, and 3) \textit{incremental mechanism design} where the design starts from some (typically well-known but not strategyproof) mechanism and then keeps making changes to the mechanism to improve it~\cite{Conitzer07Incremental}.

A recent form of incremental mechanism design that capitalizes on the modern power of deep learning is called \textit{differentiable economics}. \citet{Duetting2019Optimal} introduced the use of deep neural networks as function approximators to learn auctions. Their RegretNet architecture learns approximately strategyproof auctions for multi-bidder multi-item auctions.
MenuNet~\cite{Shen2019Automated} and RochetNet~\cite{Duetting2019Optimal} are restricted to a single bidder, but enforce strategyproofness at the architectural level.

\section{Our Contributions}
Ideally, we would like an auction architecture that 1) supports multiple agents and items, 2) is perfectly strategyproof by construction, and 3) is always rich enough to represent the true optimal auction, given enough parameters.
Such an architecture does not yet exist.
RegretNet achieves 1 and 3 only; RochetNet and MenuNet achieve 2 and 3.
In our work, we present an approach that achieves 1 and 2, though not 3 -- \textbf{a multi-bidder, multi-item auction architecture which is always perfectly strategyproof}.

Consider a classic tool for automated mechanism design -- the family of \textit{affine maximizer auctions (AMAs)}~\citep{Roberts1979Characterization}. AMAs are essentially versions of the VCG mechanism, modified by associating a positive ``weight'' to each bidder's welfare and adding potentially different ``boosts'' to all the possible allocations. AMAs are always strategyproof and individually rational like VCG, but revenue can be significantly increased over VCG by tuning these parameters (weights and boosts). Importantly, this can be done by just using \textit{samples} of the valuation distribution~\cite{Likhodedov2004Methods,Likhodedov2005Approximating,Sandholm2015Automated} rather than the traditional mechanism design approach of taking the full valuation distribution as input, which would be prohibitively complex in these combinatorial settings. Later work considers the number of samples needed for this in a learning-theoretic sense~\cite{Balcan2016Sample,Balcan2018General,Balcan2021How}.

Our contribution is to revisit the problem of learning AMAs, now with differentiable economics. One can view the paper from at least the following perspectives:
\begin{enumerate}
    \item It can be seen as an extension of previous work on learning AMAs, now \textbf{allowing for lottery allocations}. This means not only learning the weights and boosts, but also learning over the (continuous) set of lotteries to offer. Randomization can increase revenue.

    \item It can be seen as a \textbf{multi-bidder generalization of RochetNet and MenuNet}. Restricting our lottery AMAs to a single bidder essentially recovers these architectures, and for multiple bidders, strategyproofness is still guaranteed by construction. (However, for general multi-bidder combinatorial auction settings, AMAs cannot represent every truthful mechanism; there is no guarantee they can represent an optimal one.)
    
    \item It provides a \textbf{more interpretable} family of mechanisms to learn using differentiable economics. RegretNet-style auctions are opaque: they map bid profiles to outcomes in an arbitrary way. In contrast, the rules for determining outcomes of an AMA are easy to explain. Moreover, by the end of training, our learned mechanisms typically have a small number of possible outcomes which are easily summarized.
\end{enumerate}

\begin{figure*}[t]
    \centering
    \includegraphics[width=2.0\columnwidth]{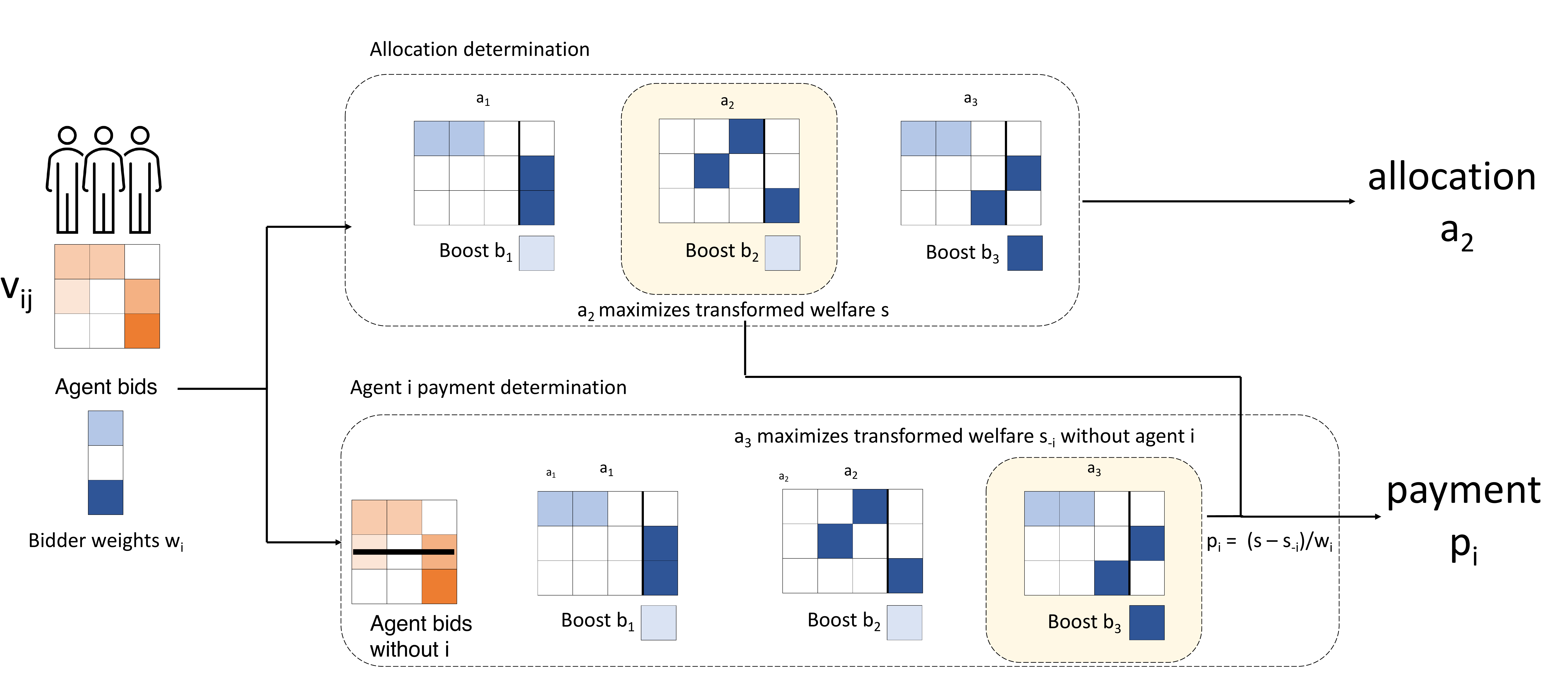}
    \caption{A schematic of an affine maximizer auction. The learned parameters---allocations, bidder weights, and allocation boosts---are represented \textcolor{powerpointblue}{in blue}. (In a traditional AMA, allocations are not learned.) The chosen allocation maximizes transformed total welfare; the payment for each bidder $i$ is the difference in transformed welfare between the chosen allocation, and what would have been chosen without taking bidder $i$ into account.}
    \label{fig:amadiagram}
\end{figure*}
\section{Related work}

\subsection{Differentiable economics}

\paragraph{RegretNet}
\cite{Duetting2019Optimal}~use the tools of modern deep learning to learn revenue-maximizing mechanisms. 
In particular, they present the RegretNet neural architecture.
The idea is to treat an auction mechanism as a function mapping bid profiles to allocations and payments, and directly approximate this function using a neural network.
The loss function consists of a term for revenue maximization, and another term for minimizing \textit{regret} -- violations of strategyproofness.
RegretNet works quite well, approximately recovering some known optimal auctions and outperforming other approaches.

However, its approach has several limitations.
In particular, the learned auctions are only approximately strategyproof -- there is still some small presence of regret, and moreover the presence of regret can only be measured empirically.
\citet{Curry2020Certifying} provides a way to exactly compute regret, which mitigates this latter limitation.
But the former problem remains -- a mechanism learned using the RegretNet approach is not guaranteed to be perfectly strategyproof.
\subsection{Characterizing strategyproof mechanisms}
\citet{Rochet1987necessary} shows that for any mechanism \textit{with a single agent}, strategyproof mechanisms can be identified with convex utility functions (as a function of the agent's true type).
Any strategyproof pair of allocation and payment rules will induce a convex utility function.
An allocation rule can be derived from any convex utility function by simply taking its gradient (which also fixes the payment rule).

Characterizing strategyproof mechanisms for multiple agents is not so straightforward.
Rochet's characterization still holds in this case: fixing other bids, agent $i$'s utility must be convex as a function of their type, and this must hold for all agents and for any choice of opponent bids.
However, coming up with some universal approximator for the entire class of functions that has this property is difficult.

\subsection{Strategyproof architectures}
Alongside RegretNet, \citet{Duetting2019Optimal} also presents the RochetNet architecture, which is restricted to a single bidder but is perfectly strategyproof.\footnote{In the appendix, they also present MyersonNet, which is restricted to 1 item.} \citet{Shen2019Automated} concurrently present MenuNet, another single-bidder architecture which is perfectly strategyproof.

Both MenuNet and RochetNet offer possibly-randomized sets of menu items at different prices.
The bidder maximizes over all offered menu items, inducing a convex utility as a function of the bidder's type.
As such, MenuNet and RochetNet will always represent a strategyproof mechanism for any setting of their parameters.
And given enough parameters, they are universal approximators for strategyproof mechanisms.

For single-bidder auction design, there is a strong duality result which can be used to prove optimality of a proposed mechanism~\cite{Daskalakis2017Strong,Kash2016Optimal}.
The authors of \citet{Duetting2019Optimal} and \citet{Shen2019Automated} apply these results to their learned auctions, and discover some previously-unknown optimal auctions.
\subsection{Further work in differentiable economics}
Many papers have built on RegretNet.
ALGNet~\cite{Rahme2021Auction} gives an improved loss function, which has fewer hyperparameters, and an improved training algorithm.
We use it as a point of comparison below.
Other papers apply the same general approach to auctions with fairness or budget constraints \cite{Kuo2020ProportionNet,Peri2021Preferencenet,Feng2018Deep}, add new inductive biases to the architecture \cite{Curry2021Learning,Rahme2020Permutation}, or apply similar techniques to other mechanism design problems \cite{Ravindranath2021Deep,Golowich2018Deep,Brero2021Reinforcement}.

Another line of work uses neural networks to model agent preferences over possible bundles~\cite{Tacchetti2019Neural,Weissteiner2020Deep,Brero2019Machine,Brero2019Fast,BachrachNeural2021}.
\citet{Bichler2021Learning} uses ML techniques to compute equilibrium strategies for non-incentive-compatible auctions.

\subsection{Automated mechanism design and learning theory for auctions} 
Affine maximizer auctions (AMAs) are classic tools for automated mechanism design~\cite{Sandholm2015Automated,Likhodedov2004Methods,Likhodedov2005Approximating}. 
In essence, AMAs are just weighted versions of the celebrated Vickrey-Clarke-Groves (VCG) mechanism~\cite{Vickrey1961Counterspeculation,Clarke1971Multipart,Groves1973Incentives}.
VCG chooses the welfare-maximizing allocation; an AMA maximizes a rescaled and shifted version of the welfare.
By choosing the parameters of the AMA carefully, performance on metrics other than welfare maximization can be improved without sacrificing strategyproofness.
Previous work considers the problem of learning high-performing AMAs from samples using gradient based methods~\cite{Sandholm2015Automated,Likhodedov2004Methods}, albeit using different techniques and without considering lotteries.
\citet{Guo2017Optimizing} computes AMA parameters via linear programming for a particular problem setting.
Other works consider the sample complexity of learning AMAs, treating them as a parameterized function class~\cite{Balcan2016Sample,Balcan2018General,Balcan2021How}.
\citet{TangMixed2012} considers a subset of AMAs for which the optimal revenue can be computed in closed form.
\citet{DengTowards2021} tunes the parameters of a class of AMAs to improve performance in an online advertising application.

\subsection{Lotteries and menu size complexity} There are a number of theoretical results showing that offering lotteries can improve revenue~\cite{Briest2010Pricing,Pavlov2011Optimal,Daskalakis2017Strong}.
\citet{Hart2019Selling} analyze this phenomenon and give an interesting perspective -- in the most general sense, it is not offering lotteries \textit{per se} that improves revenue. 
Rather, it is that offering more menu items can improve revenue by allowing finer price discrimination, and there are always fewer deterministic allocations than possible lotteries.

These results, however, give worst-case revenue gaps across whole classes of valuations, not a guarantee for any specific instance. 
As discussed below, we find that even when our mechanisms can improve their revenue by offering lotteries, they offer relatively few menu items, so performance improvements are not due to increased menu size.

\subsection{Expressiveness of AMAs and Roberts's Theorem} To what extent can the class of affine maximizer auctions actually express the optimal strategyproof auction? 
As mentioned, \citet{Rochet1987necessary} shows that all single-agent strategyproof mechanisms can be identified with convex functions.
For multi-agent multi-item settings with unrestricted valuations (meaning every agent may get any positive or negative utility from any outcome, and may even care about which particular items other agents receive), \citet{Roberts1979Characterization} shows that every strategyproof mechanism must take the form of an AMA.

The settings we consider here do not have unrestricted valuations, so Robert's theorem does not apply. 
In particular, Roberts's theorem does not hold for deterministic combinatorial auctions where valuations are monotonically increasing in receiving more items, and the empty set has zero value. 
All the valuations we consider have these properties. 
On the other hand, for many settings, \cite{Lavi2003Towards} shows that any implementable allocation rule which satisfies certain natural conditions must be ``almost'' an AMA in a certain technical sense.

\section{Affine Maximizer Auctions}

\subsection{Combinatorial Auction Setting}
Consider a setting in which $m$ auction participants are bidding on $n$ items.
Each bidder has a private type $v_i \in \mathbb{R}^n$ denoting how much they value each item.

Allocations consist of matrices $a \in \mathbb{R}^{mn}_+$, where $a_{ij}$ denotes the amount of item $j$ given to bidder $i$.
We require that $\sum_{j} a_{ij} \leq 1$, so that no item is overallocated.
For deterministic auctions, we require that $a_{ij} \in \{0,1\}$.
For unit-demand auctions, we also require that every bidder receives at most 1 item: $\sum_{i} a_{ij} \leq 1$.
Denote the set of feasible allocations for a given setting by $A \subset \mathbb{R}^{mn}$.
We will often treat $A$ as a set with elements $a_k$.
Payments $p_i$ are simply positive scalars.
Given an allocation, bidder $i$ receives utility $u_i = \sum_j a_{ij} v_{ij} - p_i$.

The regret for player $i$ under a given bid profile is defined as the difference in utility between bidding truthfully and the best strategic misreport: 
\begin{equation*}
     \text{rgt}_i(v) = \max_{b_i} u_i(b_i, v_{-i}) - u_i(v_i)
\end{equation*}
When regret is 0 for every player, and for every bid profile, the auction is dominant-strategy incentive compatible (DSIC).
In this work, all our auctions have guaranteed zero regret, but some of our baselines may have positive regret.

In addition to requiring our auctions to be DSIC, we also require individual rationality (IR) -- that is, $u_i \geq 0$ for every bidder, or equivalently, no truthful bidder will ever pay more than the value of the items they receive.

\subsection{Affine Maximizer Auction Mechanism}
Affine maximizer auctions have parameters consisting of weights $w_i$ for each bidder and boosts $b_k$ associated with each allocation $a_k$. 
Given some bids $\bm v$ for each bidder, the affine maximizer auction chooses the allocation (and boost) $a_k, b_k$ that will maximize the weighted, boosted welfare:
\begin{equation}
    k^* = \arg\max_{k} \sum_i w_i \sum_j (a_k)_{ij} \bm v_{ij} + b_k
\end{equation}
Let $a(\bm v) = a_{k^*}, b(\bm v) = b_{k^*}$.

Then, to compute a payment $p_i$ for bidder $i$, it considers the counterfactual auction result where bidder $i$ did not participate.
The total decrease in all other bidder' welfare (weighted and boosted) between this counterfactual auction and the new auction is $p_i$:
\begin{equation}
\begin{aligned}
    p_i = \frac{1}{w_i} \left(\sum_{\ell \neq i} \sum_j w_\ell a(\bm v_{-i})_{\ell j} \bm v_{\ell j} + b(\bm v_{-i})\right)\\- \frac{1}{w_i} \left(\sum_{\ell \neq i} \sum_j w_\ell a(\bm v)_{\ell j} \bm v_{\ell j} + b(\bm v) \right)
    \end{aligned}
\end{equation}
As mentioned above, AMAs (like the VCG mechanism) are always DSIC. 
To see why this is the case, observe that for any fixed set of bids $v_{-i}$, agent $i$'s utility $u_i(v_i) = \sum_j a(v_i, v_{-i})_{ij} v_{ij} - p_i(v_i, v_{-i})$ will be a pointwise maximum over a set of affine functions (one per possible allocation), and thus convex.

The choice of the above payment rule also ensures IR. We additionally require that allocating nothing and charging nothing always be among the possible outcomes $a_k$, although this is not strictly required to ensure IR.

\paragraph{Our Approach Generalizes RochetNet and MenuNet} When there is only one bidder, without loss of generality we can fix the weights to one and assume welfare when the single bidder is removed is zero, recovering the max-over-affine representation of a strategyproof single-bidder mechanism.
Thus our approach of learning allocations and boosts by gradient descent directly generalizes RochetNet \cite{Duetting2019Optimal} and MenuNet \cite{Shen2019Automated}.

\section{Learning Affine Maximizers Via Differentiable Economics}

AMAs have three types of parameters: the bidder weights $w_i$, the boosts $b_k$, and the allocations $a_k$.
(Treating the allocations of AMAs as learned parameters along with the weights and boosts is a contribution of our work.)
We assume access to sampled truthful valuations, and learn these parameters jointly via gradient descent on the objective $-\sum_i p_i$.

During training, we use the softmax function as a differentiable surrogate for the max and argmax operations: that is, 
$\arg\max_k f(a_k) \approx \langle \softmax_{\tau}(f(a_1), \cdots, f(a_k)), \bm a \rangle$
and 
$\max_k f(a_k) \approx \langle \softmax_{\tau}(f(a_1), \cdots, f(a_k)), \bm f(\bm a) \rangle$
As the softmax temperature parameter $\tau$ approaches 0, this approach recovers the true argmax.

Using this soft version of the AMA definition, we directly compute the total payment and differentiate it with respect to the parameters via the Jax autograd system~\cite{Jax2018Github} along with Haiku~\cite{haiku2020github} and optax~\cite{optax2020github}.
At test time, we use the learned parameters in the exact AMA definition, using the regular max operator.

For deterministic auctions, we fix the set $a_k$ to be the set of all feasible allocations. 
For lottery auctions, we randomly initialize a large (typically $|A| = 4096$) set of allocations -- although by the end of training, very few of these are actually used (discussed below). 

We parameterize these allocations $a_k$ to ensure that they are always feasible. 
Following the approach from \cite{Duetting2019Optimal}, for additive allocations, each allocation is represented an $m$ by $n+1$ matrix of unrestricted parameters -- the extra column is for a dummy item representing ``no allocation''.
We take an item-wise softmax and truncate the dummy column to generate a feasible allocation.
For unit-demand allocations, we follow the approach used in~\cite{Ravindranath2021Deep}, applying the softplus operation to two matrices of $m$ by $n$ parameters, normalizing row- and column-wise respectively, and taking the minimum of the result.

\section{Results}

\subsection{Hyperparameters and training} For lottery AMAs, we allow either 2048 or 4096 allocations. 
The softmax temperature is 100; we use an Adam optimizer with learning rate of $0.01$. 
We train all auctions for 9000 steps, with $2^{15}$ fresh valuation samples per gradient update.
All reported test revenues are on 100000 sampled valuations.
Because the valuation distributions are symmetric, in the cases tested below we fix bidder weights to 1.

To determine which allocations are actually used, we sample 100000 test valuations, and include any allocation that was chosen for even one bid profile.

For baselines, we compare against previously reported results from RegretNet \cite{Duetting2019Optimal}, ALGNet \cite{Rahme2021Auction},  and AMAs trained using other methods \cite{Sandholm2015Automated}, as well as theoretical revenues from Myerson auctions of separate items and of the grand bundle.

\begin{table*}[t]
\centering
\begin{tabular}{@{}lrrrr@{}}
\toprule
AMA Type      & \multicolumn{1}{l}{Max Revenue} & \multicolumn{1}{l}{Min Revenue} & \multicolumn{1}{l}{Mean Revenue} & \multicolumn{1}{l}{Std Revenue} \\ \midrule
Lottery       & 2.158                          & 1.87                           & 2.06                             & 0.098                          \\
Deterministic & 1.462                          & 0.627                          & 0.842                           & 0.279                          \\ \bottomrule
\end{tabular}
\caption{Results from 8 random parameter initializations, with 2048 allocations, on the spherical valuation distribution. In particular, the worst lottery mechanism outperforms the best deterministic mechanism. Moreover, for this setting, the lottery mechanisms do actually learn to randomize.}
\label{tab:sphericaltable}
\end{table*}

\subsection{Revenue performance}

\paragraph{Spherical distribution} In order to demonstrate a revenue improvement by offering lotteries, we consider a particular valuation distribution which we refer to as the ``spherical distribution'' for lack of a better name -- this is a distribution on a number of discrete, random points, scaled and normalized according to the proof construction in \cite{Briest2010Pricing}. 

We construct such a distribution for 4 items with 5 valuation points and consider a setting with two unit-demand bidders, each of whose valuations are sampled i.i.d from this distribution. 
We would expect a large gap between revenue extracted by lotteries and by a deterministic mechanism.

\begin{figure}
    \centering
    \includegraphics[width=\columnwidth]{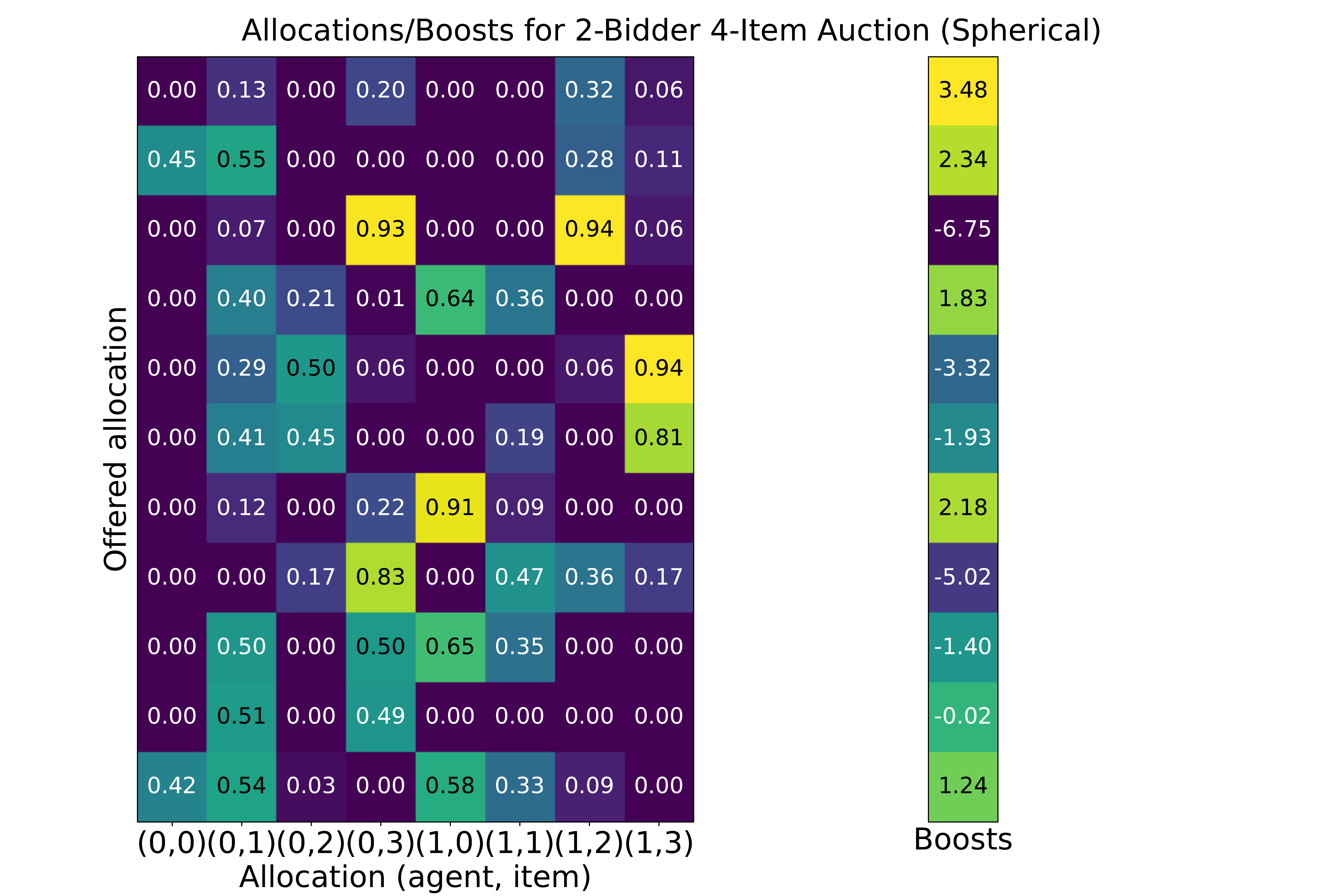}
    \caption{The ten lottery allocations (and their boosts) actually used after training an auction. (The auction has many more parameters, but the 2038 other allocations are never chosen for any of the sampled bids.) One can see that the mechanism does typically offer lotteries.}
    \label{fig:2x4spherical}
\end{figure}

Indeed, we find that this is the case -- when we train our lottery AMA with 2048 allocations on this distribution, it gets more than twice the revenue of a deterministic AMA (see \ref{tab:sphericaltable}). 
Figure \ref{fig:2x4spherical} shows the final offered allocations and boosts from a representative mechanism -- the auction is actually taking advantage of randomization.

\paragraph{2 bidder, 2 item uniform} We also consider a 2 bidder, 2 item additive auction where item values are independently distributed on $U[0,1]$. This seems like the most trivial possible multi-bidder multi-item auction setting, yet it is so far completely beyond current theory -- this makes it an interesting test case for automated mechanism design. 
\begin{figure}[t]
    \centering
    \includegraphics[width=\columnwidth]{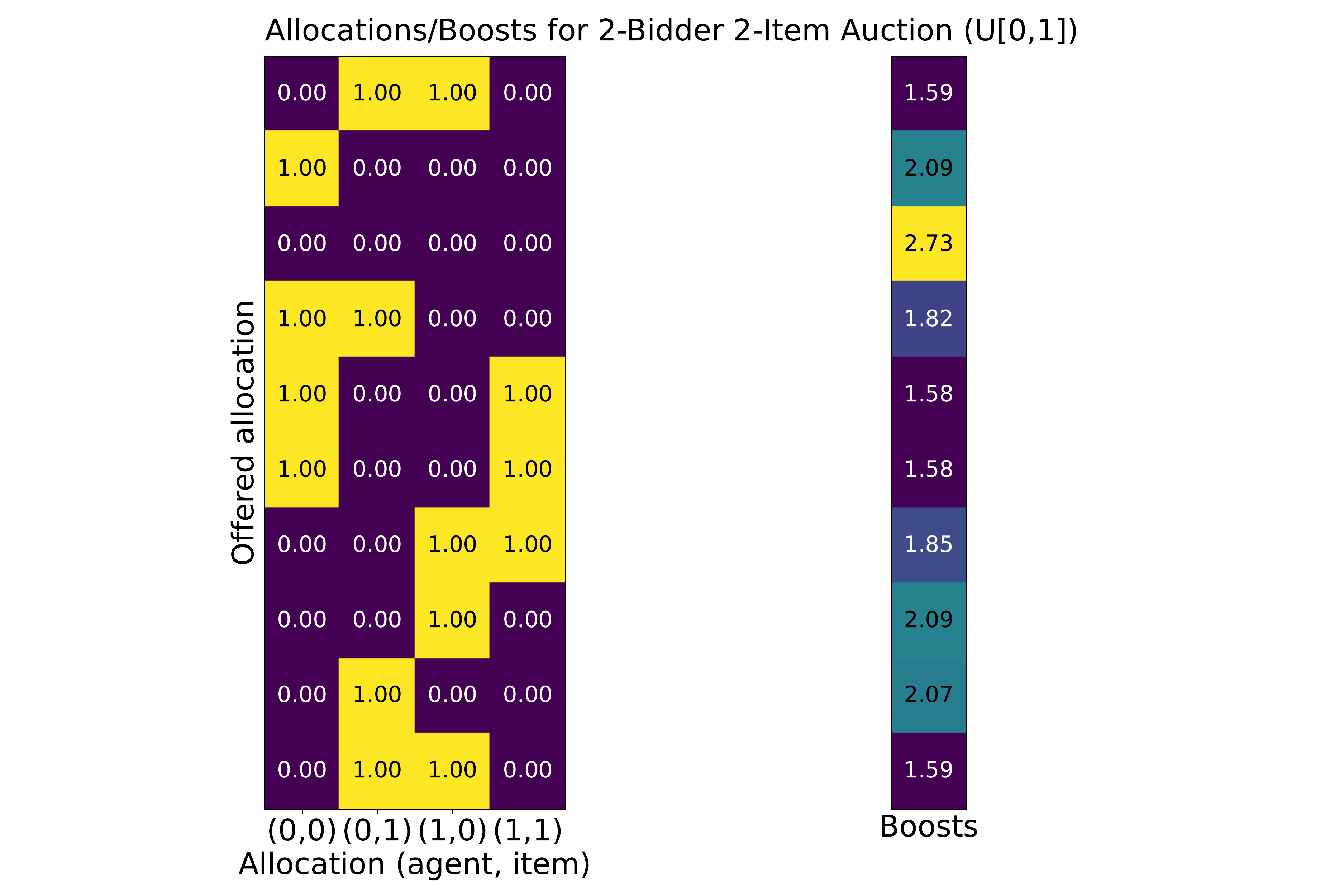}
    \caption{All allocations actually used after training for a 2x2 U[0,1] additive auction (the same setting as compared to in \citet{Likhodedov2004Methods} and \citet{Duetting2019Optimal}). Here, although the mechanism space is that of randomized mechanisms, the algorithm learns to offer deterministic allocations. The revenue is comparable to results in \citet{Likhodedov2004Methods}.}
    \label{fig:mvlearned}
\end{figure}

We train a lottery AMA on this setting and find revenue competitive with both previous AMA approaches \cite{Sandholm2015Automated,TangMixed2012} as well as the RegretNet neural network approach (which performs better but is not perfectly strategyproof) \cite{Duetting2019Optimal}.

An interesting observation, though, is that even though our lottery AMA is free to offer lotteries, it does not do so -- all allocations actually offered by the end of training are deterministic, as seen in Figure \ref{fig:mvlearned}.
\begin{table}[]
\centering
\begin{tabular}{@{}lrl@{}}
\toprule
Auction        & \multicolumn{1}{l}{Best Revenue} & Regret                \\ \midrule
Lottery AMA (ours)   & 0.868                           & 0 \\
Combinatorial AMA & 0.862                            & 0 \\
Separate Myerson & 0.833 & 0 \\
Grand Bundle & 0.839 & 0 \\
MBARP & 0.871 & 0 \\
RegretNet      & 0.878                            & $<0.001$     \\ 
ALGNet & 0.879 & 0.00058 \\\bottomrule
\end{tabular}
\caption{Revenue comparison for 2 bidder, 2 item U[0,1] additive auction. Our approach is competitive with other approaches. Combinatorial AMA refers to results from \citet{Sandholm2015Automated}. MBARP is a subset of AMA from \citet{TangMixed2012} where the optimal parameters have been computed (only for 2 items). RegretNet achieves higher revenue, but possibly due to a small strategyproofness violation. Note that \citet{Sandholm2015Automated} present many variants, some of which beat our revenue, although all are comparable.}
\label{tab:mvtable}
\end{table}

\paragraph{3 bidder, 10 item uniform} Finally, we consider one of the much larger auction settings from \cite{Duetting2019Optimal} -- 3 additive bidders with 10 $U[0,1]$ items. We give our network parameters for 4096 allocations, many fewer than the number of possible deterministic allocations in this setting.

Results are shown in Table \ref{tab:3x10}, along with baselines. 
While we do not match the performance of RegretNet and ALGNet, we do at least exceed the performance of the separate Myerson and grand bundling approaches.
Some, though probably not most, of the extra revenue gained by RegretNet and ALGNet may be due to non-zero regret.

\begin{table}[]
\centering
\begin{tabular}{@{}lrr@{}}
\toprule
Auction          & \multicolumn{1}{l}{Best Revenue} & \multicolumn{1}{l}{Regret} \\ \midrule
Lottery AMA (ours)     & 5.345                         & 0                          \\
Separate Myerson & 5.31                             & 0                          \\
Grand bundle     & 5.009                            & 0                          \\
RegretNet        & 5.541                            & 0.002                      \\ 
ALGNet        & 5.562                           & 0.002                      \\ \bottomrule

\end{tabular}
\caption{Revenue comparison for 3 bidder, 10 item U[0,1] additive auction. We train a lottery AMA with 4096 allocations. It underperforms RegretNet (although RegretNet has a small strategyproofness violation), but outperforms the separate Myerson and grand bundling baselines.}
\label{tab:3x10}
\end{table}

We also attempted to train a lottery AMA for the 5 bidder, 10 item uniform case, but found that after several attempts it failed to outperform the separate Myerson baseline.

\subsection{Number of allocations used}
\begin{table*}[]
\centering
\begin{tabular}{@{}lrrrr@{}}
\toprule
Auction          & \multicolumn{1}{l}{Min} & \multicolumn{1}{l}{Max} & \# at Initialization & \# Deterministic \\ \midrule
Lottery spherical      & 8                         & 15       & 2048 & 20                  \\
Deterministic spherical & 6                             & 9 & 20        & 20                   \\
2x2 U[0,1]     & 7                            & 10 & 4096     & 81                    \\
3x10 U[0,1]        & 58                            & 64  & 4096   &       $2^{20}$          \\ \bottomrule
\end{tabular}
\caption{The number of allocations actually used after 9000 steps of training, for the experiments given above. These quantities are smaller than the number of initial outcomes as well as the number of possible deterministic outcomes.}
\label{tab:numalloc}
\end{table*}
We observe that although our auctions are initialized with many parameters, the number of possible deterministic outcomes may be quite large, the number of allocations used on any actual valuation profile is typically quite small.
Results are summarized in Table \ref{tab:numalloc} for all experiments mentioned above.

\subsection{Effects of parameter initialization}
\label{sec:lotteryticket}
\begin{table}[]
\centering
\begin{tabular}{@{}lrr@{}}
\toprule
               & \multicolumn{1}{l}{Mean Rev.} & \multicolumn{1}{l}{Best Rev.} \\ \midrule
Winning Ticket (2x2) & 0.870                      & 0.872                   \\
Small Random (2x2)        & 0.772                        & 0.777                        \\ 
Winning Ticket (Spherical)        & 1.836                       & 1.842                       \\ 

Small Random (Spherical)        & 1.197                        & 1.572                        \\ 
\bottomrule
\end{tabular}
\caption{We take the actually-used allocations from the best-performing 2x2 uniform and spherical models -- the values of these parameters \textit{before training} are the ``winning ticket''. We initialize a lottery AMA using the winning ticket initializations, and train on 4 random data seeds. To compare, we also test 4 different random initializations of the same small number of allocations, and find significantly lower performance.}
    \label{tab:lottery2}
\end{table}

Motivated by the \textit{lottery ticket hypothesis} in neural network training~\cite{Frankle2018Lottery}, we consider the effects of overparameterization and parameter initialization on performance.

First, we consider training from the same parameter initialization, under a different source of randomness for the data.
We find that starting from the same initialization typically results in nearly the same allocation indices being chosen, with Jaccard similarities of .64, .67, and .82 across the indices chosen under the new random data.
Starting from a different parameter initialization, there was no overlap in the indices chosen.
We find that the results are quite similar, which suggests that parameter initialization is important in determining the end results.

We also consider the opposite approach: take the final actually-used allocations, look at what values those parameters took at initialization before training, and retrain using only those parameters.
In other words, we train a model with very few parameters ``from scratch'', but with an initialization we hope will perform well -- the ``winning lottery ticket''.

Results are summarized in Table~\ref{tab:lottery2}.
We indeed find a large gap in performance between the good initialization and randomly-initialized models with the same number of parameters.

\section{Discussion}

We see our approach as a first step towards strategyproof architectures for multi-agent differentiable economics.
On the one hand, it is a natural generalization of RochetNet and MenuNet. 
On the other hand, it is also a natural generalization of classic work on AMAs.

Beyond the obvious advantage of perfect strategyproofness, there are other reasons one might prefer this approach over RegretNet. In particular, AMAs are interpretable -- it's easy to simply inspect which allocations are being offered as possibilities.
However, it's unclear when and whether our approach can actually represent the true optimal mechanism -- that remains an open theory question.
Regardless, we see it as a useful tool for automated mechanism design in multi-bidder multi-item settings.

\subsection{Lottery ticket hypothesis} Our networks are quite sensitive to initialization -- there's a relatively wide range of performance between reinitialized instances of the same architecture shown the same sequence of training data.
Moreover, we found that starting out with a large number of parameters improves performance, even though by the end of training only a tiny number of these parameters were actually used.

A dependence on initialization, a benefit from overparameterization, and a final model which is effectively sparse all bring to mind the lottery ticket hypothesis~\cite{Frankle2018Lottery} in deep learning.
Indeed, our experimental results in section \ref{sec:lotteryticket} suggest that some version of the lottery ticket hypothesis is in play here.
We also observe that \citet{Curry2020Certifying} was able to significantly distill learned auction networks without harming performance.
Future work in auction learning might further take advantage of this direction.

\subsection{Strengths and Limitations}
\paragraph{Limitations compared to other differentiable economics approaches} The most obvious limitation of our work is that in most settings, there are probably strategyproof non-AMAs which outperform the best AMA, but we cannot learn these.
This is most likely why ALGNet and RegretNet outperform our mechanisms in terms of revenue.
On the other hand, we think that ensuring perfect strategyproofness is a real advantage, and a more flexible neural architecture that preserves this property while going beyond AMAs remains out of reach.
\paragraph{Limitations compared to other AMA/AMD approaches}
Beyond this, our approach works for settings with additive and unit-demand valuations, where the bidders' valuations are just vectors so that the total value of a bundle can be expressed as an inner product.
It is not straightforwardly well-suited to complex combinatorial valuations where bidders may value bundles very differently depending on the presence of specific items.

Moreover, we solve the winner determination problem by explicitly computing the (weighted, boosted) welfare for every possible allocation.
This works because at any given time, our auctions are only offering a restricted set of learned allocations, which is usually much smaller than the overall set of possible allocations.
But in a setting where it was important to offer something closer to the full set of allocations, our approach would quickly become intractable.

\paragraph{Strength: interpretability}
A major advantage of our approach is in interpretability.
Neural-network-based approaches are almost totally opaque -- there's no way to summarize or explain the learned mechanism to the bidders, other than simply giving them the network parameters.
By contrast, in our approach, we can simply describe the possible allocations, their boosts, and each bidder's weights, and it is immediately clear how the mechanism works.
\section{Future Research}

We focus on auctions because there is a large body of techniques in automated mechanism design and differentiable economics which provide useful baselines for performance.
But VCG-style mechanisms can be used for mechanism design problems beyond auctions, or for more complex types of auctions than considered here.

We expect that the approach described here could be extended to other mechanism design problems as long as 1) feasible mechanism outcomes can be parameterized in a way amenable to gradient-based learning, and 2) the welfare of an outcome as a function of agent types can be computed in a way that preserves differentiability.
Exploring the use of learned AMAs in new mechanism design settings is a fruitful direction for future work.

\section{Acknowledgements}
Curry and Dickerson were supported in part by NSF CAREER Award IIS-1846237, NSF D-ISN Award \#2039862, NSF Award CCF-1852352, NIH R01 Award NLM-013039-01, NIST MSE Award \#20126334, DARPA GARD \#HR00112020007, DoD WHS Award \#HQ003420F0035, ARPA-E Award \#4334192, and a Google Faculty Research Award. Sandholm was supported by the National Science Foundation under grants IIS-1901403 and CCF-1733556.
We thank Ping-yeh Chiang, Jonas Geiping, Uro Lyi, David Miller, and Neehar Peri for helpful comments and guidance on early versions of this work.
\bibliography{references}
\bibliographystyle{icml2022}
\end{document}